\documentclass[12pt,thmsa]{article}

\begin{document}

\author{John F. Dobson and Evan MacA. Gray \\
Nanoscale Science and Technology Centre, Griffith University, \\
 Nathan, Queensland 4111, Australia}
\title{Residual attractive force between superparamagnetic nanoparticles}
\date{\today}
\maketitle

\begin{abstract}
A superparamagnetic nanoparticle (SPN) is a nanometre-sized piece of a
material that would, in bulk, be a permanent magnet. In the SPN the
individual atomic spins are aligned via Pauli effects into a single giant
moment that has easy orientations set by shape or magnetocrystalline anisotropy.  
Above a size-dependent blocking temperature $T_{b}(V,\tau _{obs})$ , thermal
fluctuations destroy the average moment by flipping the 
giant spin between easy orientations at a rate that is rapid on the scale of the observation time $\tau _{obs}$.
 We show that, depite the vanising of the average moment,
two SPNs experience a net attractive force of magnetic origin, analogous to the van der Waals force
 between molecules that lack a permanent electric dipole. This could be
relevant for ferrofluids, for the clumping of SPNs used for drug delivery,
and for ultra-dense magnetic recording media.
\end{abstract}


\section{Introduction}

In many areas of physics, forces are effectively suppressed in the
interaction between separated fragments of matter, because of the neutrality
of each fragment with respect to the appropriate charge quantity.
Nevertheless ''residual'' forces still occur between these fragments,
typically with a decay (as a function of the spatial separation $D$ between
the fragments) that is different from that of the ''bare'' interaction.

For example, ordinary matter consisting of atoms and molecules is typically
neutral with respect to electrical charge, but two well-separated
charge-neutral fragments always experience at least the van der Waals or
dispersion interaction. This is a residual force that arises because the
zero-point motions of the electrons on the two fragments are correlated via
the Coulomb interaction, leading to a non-zero time-averaged force of
Coulombic origin, despite overall charge neutrality of each fragment. For
neutral molecules distant $D$, this leads to an interaction energy varying
as $-D^{-6}$. This is to be compared with the bare Coulomb interaction
proportional to $Q_{1}Q_{2}D^{-1}$ that acts between between fragments with
nonzero electric charges $Q_{1}$, $Q_{2}$. \ ($D^{-6}$ is replaced by $%
D^{-7} $ when $D$ is large enough that retardation of the electromagnetic
interaction needs to be considered \cite{MahNin,ParsegianBook}).  

Similarly the nuclear force between two nucleons has sometines
been regarded as a residual color interaction between color-neutral objects.

Here we propose a similar residual force, of magnetic dipolar origin, acting
between two ''superparamagnetic nanoparticles (SPNs)''. By this we mean that
each nanometre-sized particle is composed of a material that is
ferromagnetic in its bulk state 
\cite{ThFluctSPNFullerBrown,MagneticNanoparticlesGenDrugDel}. 
Typically at the temperatures of
interest, the elementary electron spins inside an individual nanoparticle
remain locked together by the microscopic exchange interaction, yielding
effectively a single giant spin with a magnetic dipole moment $d_{0}$. If
the directions of the giant moments remain steady over time, two such
nanoparticles experience a conventional magnetic dipole-dipole energy
proportional to $d_{0}^{(1)}d_{0}^{(2)}f(\theta _{1},\phi _{1};\theta
_{2},\phi _{2})/R^{3}$. Here $R$ is the spatial separation of the
nanoparticles, and $f$ is a dimensionless function of the angles between
each fixed moment and the vector $\vec{R}$ joining the spatial locations of the
nanoparticles. However each particle has one or more "easy axes" in directions 
determined by magnetocrystalline or, more typically, shape anisotropy.
 The latter effect arises in the strong angular dependence of the magnetostatic 
self-energy
 of a non-spherical magnetised particle. We will consider the simplest case, 
in which the particle is sufficiently elongated that it has a single easy axis,
 i.e. dominant uniaxial shape anisotropy. Then the energy of a single 
nanoparticle is lowest when its giant spin (dipole moment, $\vec{d }$) lies
 parallel or antiparallel to this easy axis. Because the energy barrier
 $E_0$ for rotation of $\vec{d}$  between easy orientations 
(not mechanical rotation of the particle) derives from the magnetic self-energy
 of the nanoparticle, it decreases with decreasing volume of the nanoparticle. 
For very small particles, therefore, the projection of  $\vec{d}$ on a measurement
 axis averages to zero over time, because of repeated thermal flipping of the 
giant spin \cite{ThFluctSPNFullerBrown}, caused by thermal agitation from the
 heat bath (e.g. a fluid or solid matrix) that surrounds the nanoparticle. Thus
 on time average the nanoparticle is "neutral" i.e. it has  a zero magnetic moment.

When the thermal agitation of the giant spin is insufficient to flip it between easy
 orientations within the observation time, $\tau_{obs}$, the nanoparticle is "blocked",
 i.e. apparently frozen as to its magnetism. This occurs below the blocking 
temperature of this nanoparticle, $T_b$, which depends on $E_0$ and therefore 
on the volume $V$ of the nanoparticle. If the relaxation time of $\vec{d}$ over the
 barrier $E_0$ is $\tau$, then $T_b$  is defined by $\tau(T_b) = \tau_{obs}$. 
Blocking is thus a purely dynamic phenomenon: extending the observation time, or 
lowing the frequency, lowers $T_b$ and vice-versa  \cite{ThFluctSPNFullerBrown}.
  For the present case of SPNs suspended in a fluid, the observation 
time $\tau_{obs}$ will be a relevant time for mechanical motion of the SPN
 through the fluid. - e.g. a rotational or translational diffusion time. 
 Note that the direct dipolar magnetic interaction between SPNs could in principle
 lead them to clump. However when $T > T_b$ the motion of the SPNs through the 
fluid will not "see" the bare dipolar magnetic interaction between the SPNs, 
as it has been averaged away between attractive and repulsive values during
 the thermal flipping of the spins.  It could lead to additional Brownian type 
of damping and difusion of course, but we show here that there is also a net 
attractive force between SPNs even above the blocking temperature.

The destruction of the permanent magnetic moments by thermal fluctations is highly
undesirable in the case of a magnetic data recording medium, where very fine
magnetic particles in the nanometer size range will be needed in order to
pack the magnetically stored data as densely as possible for the next
generation of devices. The thermal destruction of the permanent moments means that
data cannot be stored over long times.

On the other hand, as will be discussed below, the same thermal flipping
occuring for $T> T_{b}$ is beneficial in the case of nanoparticles
deliberately suspended in human blood as carriers for drug or thermal
therapies, since now the clumping of the nanoparticles from magnetic dipole
interaction is suppressed because each particle has effectively a zero
magnetic moment. \ The strong clumping that would occur for fully
ferromagnetic particles from their $R^{-3}$ dipole-dipole interactions could
be clinically dangerous, potentially causing blockage of blood vessels,
difficulty of elimination etc. \ We will show below, however, that despite
the vanishing of the average individual moments, there is a residual attractive
interaction between two superparamagnetic nanoparticles separated by
distance $R,$ that falls off as $(const)/R^{6}$. It is the magnetic analog
of the van der Waals or dispersion force that arises via the Coulomb
interaction between fluctuating \emph{electric} dipoles on two electrically
neutral molecules \cite{SoftCohForces} lacking permanent dipole moments. 
 This residual force could also lead to clumping of the
nanoparticles, and so its analysis could be signifiant in modern
magnetic-particle therapies \cite{MagneticNanoparticlesGenDrugDel}.

\section{Simple preliminary model}

The model described here is based on an argument frequently used to explain
the attractive van der Waals energy proportional to $-R^{-6}$ that arises
between temporary electric dipoles occurring on a pair of electronically
neutral atoms separated by distance $R$ (see e.g. \cite{SoftCohForces}). It is not rigorous derivation, but
may help to elucidate the more careful and general mathematical treatment to
be provided in later Sections. Consider two superparamagnetic nanoparticles
SPN1 and SPN2 as defined above. While averaging to zero over time as
described above, the magnetic moment $\vec{d}^{(1)}$ on SPN1 can exhibit a
short-lived thermal (or quantal) fluctuation so that its value $\vec{d}%
^{(1)}(t)$ is nonzero at some particular time $t$. \ For simplicity we will
assume that only magnetizations of SPN1 and SPN2 along one axis (say $\hat{z}$%
) are possible so that $\vec{d}^{(1)}(t)=d^{(1)}(t)\hat{z}$, and we will
consider the case that the spatial separation $\vec{R}$ between SPN1 and
SPN2 is parallel to $\hat{x}$. Then the spontaneous moment $\vec{d}_{1}(t)$
produces a dipolar magnetic induction (B-field)%
\[
\vec{b}^{(2)}(t)=-\frac{\mu _{0}}{4\pi }R^{-3}d^{(1)}(t)\hat{z} 
\]%
at the position of SPN2. Responding to this field, SPN2 produces its own
magnetic moment 
\begin{equation}
\vec{d}^{(2)}(t)=\bar{\chi} ^{(2)}\vec{b}^{(2)}(t)=-\chi ^{(2)}\frac{\mu _{0}}{4\pi }%
R^{-3}d^{(1)}(t)\hat{z}
\label{DipoleOn2} 
\end{equation}
where $\bar{\chi} ^{(2)}$ is the dynamic magnetic susceptibility of SPN2, assumed 
for now to represent an instantaneous response to the field. 
(Note that here $\bar{\chi}$ represents the response of the total magnetic moment of the SPN to a small applied magnetic induction $\vec{b}$. By contrast, the symbol $\chi$ is normally used for the response of the magnetic moment \emph{per unit volume} to a small applied \emph{magnetic field} $\vec{h}$.  Thus for a single SPN of volume $V$,
\begin{equation}
d = \bar{\chi} b;\;\;\;\;\frac{d}{V} = \chi h; \;\;\;\chi= \frac{\mu_0 \bar{\chi}}{V}
\label{ChibarVsChi}
\end{equation}

The dipole (\ref{DipoleOn2}) in turn
 produces a dipolar magnetic induction back at the position of SPN1:%
\begin{eqnarray*}
\vec{b}^{(1)}(t) &=&-\frac{\mu _{0}}{4\pi }R^{-3}d^{(2)}(t)\hat{z}=\left( -\frac{%
\mu _{0}}{4\pi }R^{-3}\right) \left( -\chi^{(2)}\frac{\mu _{0}}{4\pi }%
R^{-3}d^{(1)}(t)\right) \hat{z} \\
&=&\left( \frac{\mu _{0}}{4\pi }\right) ^{2}R^{-6}\chi ^{(2)}d^{(1)}(t)\hat{z}.
\end{eqnarray*}%
The interaction energy of this back-field with the original moment $\vec{d}%
^{(1)}(t)$ is 
\[
E=-\vec{b}^{(1)}(t).\vec{d}^{(1)}(t) 
\]%
and this energy has a time or thermal ensemble average 
\[
\left\langle E\right\rangle =-\left( \frac{\mu _{0}}{4\pi }\right)
^{2}R^{-6}\chi _{2}\left\langle (d^{(1)} (t))^2\right\rangle 
\]%
which is non-zero because $\left\langle (d^{(1)}(t))^2 \right\rangle \neq 0$
even though $\left\langle d^{(1)}(t)\right\rangle =0$.

This negative energy produces, upon differentiation with respect to $R$, a
net time-averaged attractive force between SPN1 and SPN2 that falls off as $%
R^{-7}$.

The above simplified theory produces the basic physics and the $R^{-7}$
force, but it glosses over a number of issues, such as the role of entropic
effects at finite temperature, the tensor nature of the magnetic
dipole-dipole interaction, the quantal aspects of the problem, and the
retardation of the electromagnetic field. Also, the response $\chi _{2}$ has been assumed to be instantaneous, whereas there can be a strong and important frequency
dependence (time-delayed aspect) to the linear response of a SPN. \ All of
these considerations are treated in detail in the theory given the next Section.

\section{Detailed theory}

The magnetic dipolar energy (hamiltonian) between two particles with
magnetic dipoles $\vec{d}^{(1)}$ and $\vec{d}^{(2)}$, separated in space by
a nonzero vector $\vec{R} = R\hat{R}$, is of form 
\begin{equation}
H^{(12)}=-R^{-3}\sum_{ij=1}^{3}d_{i}^{(1)}T_{ij}(\hat{R})d_{j}^{(2)}
\label{InteractionBetwMagnets}
\end{equation}%

where 
\[
T_{ij}=\frac{\mu _{0}}{4\pi }\frac{3R_{i}R_{j}-\delta
_{ij}R^{2}}{R^{2}}.
\]

We assume that we are above the blocking temperature, $T>T_{b}$, i.e. that
the temperature is high enough (compared with the anisotropy energy
barrier), that each isolated giant magnetic dipole has zero thermal
expectation taken over the time-scale of interest
\[
<\vec{d}^{(1)}>_{0}\,=\,\vec{0}=\;<\vec{d}^{(2)}>_{0}. 
\]%
The theory to be developed here is meaningful provided that the thermal fluctuations of the moment
occur on a time-scale $\tau $ that is short compared to the time $\tau_{obs} \equiv T_{mech}\,$
for the nanoparticle to change its spatial position (or physical angular
orientation) appreciably, within its fluid medium. \ Under these conditions
we will derive a residual attractive force between the two superparamagnetic
nanoparticles, that could for example be used to study residual clumping
effects in fluid suspension at temperatures above the blocking temperature .

The quantum-thermal expectation, denoted $<\;>$, of the interaction energy between the giant spins is 
\[
E^{(12)}\equiv \;<H^{(12)}>\;=-R^{-3}\sum_{ij=1}^{3}T_{ij}(\hat{R}%
)<d_{i}^{(1)}d_{j}^{(2)}>
\]%
However at finite temperature it is not this energy but the corresponding
thermal Helmholz free energy 
\[
A=\;<H^{(12)}>-TS
\]%
that must be considered, where $S$ is the entropy. \ We achieve an
expression for $A$ via a Feynman-theorem argument in Appendix A for a
classical treatment of the fluctuations, and in Appendix B for the fully
quantal case. In either case the result is 
\begin{equation}
A(\lambda =1,T,R)-A(\lambda =0,T,R)=\int_{0}^{1}E_{\lambda }^{(12)}
\frac{d\lambda}{\lambda} 
\label{FreeEnAsLambdaIntegral_1}
\end{equation}%
Here the subscript $\lambda $ means that the quantity is evaluated in the
thermal ensemble with modified interaction 
\begin{eqnarray}
H_{\lambda }^{(12)}=\lambda H^{(12)}=-\lambda
R^{-3}\sum_{ij=1}^{3}d_{i}^{(1)}T_{ij}(\hat{R})d_{j}^{(2)},\;\;0\leq \lambda
\leq 1 \\ 
E^{(12)}_{\lambda}=<H^{(12)}_{\lambda} >_{\lambda} .
\label{E12Lambda}
\end{eqnarray}
Since the coupling will be zero (equivalent to $\lambda =0$) at infinite
separation $R\rightarrow \infty $, we can write Eq (\ref{FreeEnAsLambdaIntegral_1})
as an expression for just the free energy of interaction between the two
nanoparticles: 
\begin{eqnarray}
A(T,R)-A(T,R\rightarrow \infty ) &=&\int_{0}^{1}E_{\lambda }^{(12)}\frac{%
d\lambda }{\lambda }  \nonumber \\
&=&-R^{-3}\sum_{ij=1}^{3}T_{ij}(\hat{R})\int
<d_{i}^{(1)}d_{j}^{(2)}>_{\lambda }d\lambda 
\label{FreeEnAsLambdaIntegral_2}
\end{eqnarray}%
The problem now reduces to the calculation of the equal-time
cross-correlation function $<d_{i}^{(1)}d_{j}^{(2)}>_{\lambda}$ between the moments
in a thermal ensemble with $\lambda $-reduced interaction.

The equal-time correlation function $\;<d_{i}^{(1)}d_{j}^{(2)}>_{\lambda }$ 
can be recovered from the time Fourier transform 
\[
g_{\lambda }(\omega )=\int_{-\infty }^{\infty }G_{\lambda }(t)\exp (-i\omega
t)dt=\;\int_{-\infty }^{\infty }<d_{i}^{(1)}(0)d_{j}^{(2)}(t)>_{\lambda
}\exp (-i\omega t)dt 
\]%
of the time-displaced correlation function $G_{\lambda }$, 
\[
G_{\lambda }(t)\equiv \;<d_{i}^{(1)}(0)d_{j}^{(2)}(t)>_{\lambda \;\;}. 
\]

We can use the finite-temperature fluctuation-dissipation theorem (see e.g.
\cite{LandauLifshitzGenFluctDiss}) to relate the fluctuation quantity $%
<d_{i}^{(1)}d_{j}^{(2)}>$ to the dipole-dipole response function $\bar{\chi}
_{ij}^{(12)}$ of the combined interacting system, defined in Eq (\ref%
{DefCrossChi}) below: 
\[
<d_{i}^{(1)}d_{j}^{(2)}>_{\lambda ,\omega
}\,+\;<d_{j}^{(2)}d_{i}^{(1)}>_{\lambda ,\omega }=\frac{2\hbar }{1 - \exp
(\beta \hbar \omega )}Im\left\{ \bar{\chi} _{ij,\lambda }^{(12)}(\omega
+i0)+\bar{\chi} _{ji,\lambda }^{(21)}(\omega +i0)\right\}
\]
where $\beta =\frac{1}{k_{B}T}$. 
Then 
\begin{eqnarray*}
<d_{i}^{(1)}d_{j}^{(2)}&+&d_{j}^{(2)}d_{i}^{(1)}>_{\lambda ,equal\,\,time} =%
\frac{1}{2\pi }\int_{-\infty }^{\infty }g_{\lambda }(\omega )\exp (i\omega t)
\\
&=&\int_{-\infty }^{\infty }\frac{\hbar }{\pi (1-\exp (\beta
\hbar \omega ))}Im\left\{ \bar{\chi} _{ij,\lambda }^{(12)}(\omega
+i0)+\bar{\chi} _{ji,\lambda }^{(21)}(\omega +i0)\right\} d\omega 
\end{eqnarray*}

This can also be expressed as a Matsubara sum by closing upwards in the
complex $\omega $ plane, using Cauchy's theorem to obtain a sum of residues
at the poles $\omega _{n}=iu_{n}=i2n\pi /(\beta \hbar)$, but we will not
make explicit use of this here.

The interaction energy $E$ and Helmholtz free energy $A$ then become 
\begin{eqnarray}
E_{\lambda }^{(12)}\; =-\lambda R^{-3}\sum_{ij=1}^{3}T_{ij}(\hat{R})
\int_{-\infty }^{\infty }\frac{\hbar }{\pi (1-\exp (\beta \hbar \omega))} 
Im\left\{ \bar{\chi} _{ij:\lambda}^{(12)}(\omega +i0)\right\} d\omega 
\label{IntEFromChi} \\
A(R)-A(\infty ) =\int_{0}^{1}E_{\lambda }^{(12)}\;\frac{d\lambda }{\lambda} 
  \label{HelmholtzFromChi}
\end{eqnarray}

We assume we know the dipole responses $\bar{\chi} _{ij,\lambda }^{(1)}(\omega
),\,\,\bar{\chi} _{ij,\lambda }^{(2)}(\omega )\,$ of each isolated giant dipole
SPN1, SPN2 to an external B field $\vec{b}$ such that 
\[d_{i}^{(1)}(\omega )\exp
(-i\omega t)=\sum_{j}\bar{\chi} _{ij}^{(1)}(\omega )b^{(1)}\exp (-i\omega t).
\]
These individual responses must express the known superparamagnetic
properties of individual systems. In general it should also describe any
Brownian tumbling aspects of the response, in the case that the time scale
of these tumbling motions overlaps that of the magnetic reponse behaviour of
each SPN. \ For now we assume that the tumbling is slow so that only the
magnetic response of a SPN oriented in a fixed spatial orientation is
required to appear in $\bar{\chi} $. (The interaction energy may of course depend
on the details of this orientation, which will be manifested in the
particular values of $\bar{\chi} _{ij}^{(12)}$ in the chosen cartesian frame.) In
Appendix C we discuss a simple model for the $\bar{\chi} $ of a single isolated
SPN. However, to calculate the interaction of two SPNs, Eq (\ref{IntEFromChi}%
) requres knowledge of the cross-response function (cross-susceptibility) $%
\bar{\chi}_{\lambda}^{(12)}$ for the interacting pair of SPNs. This is defined as the
linear response of SPN1's moment to an alternating B field that acts
upon SPN2 only: 
\begin{equation}
d_{i}^{1}(\omega )=\bar{\chi} _{ij,\lambda }^{(12)}(\omega )b_{j}^{(2)}\;\;,
\label{DefCrossChi}
\end{equation}%
where the subscripts $i,j$ label caretesian components of the vectors. \
To calculate $\bar{\chi} ^{(12)}\,$we now consider the slightly more general
situation where independently-specified small external B fields $\vec{b}%
^{(1)}\exp (-i\omega t),\,\vec{b}^{(2)}\exp (-i\omega t)$ are applied to the
individual dipoles, in the presence of the dipolar coupling between the two
systems.

In time-dependent mean-field theory (RPA), the equations of motion of the
coupled systems are (all at arbitrary frequency $\omega $ and with Einstein summation convention for repeated indices): 
\begin{eqnarray}
d_{\alpha }^{(1)} &=&\bar{\chi} _{\alpha \mu }^{(1)}\left( b_{\mu }^{(1)}+\lambda
R^{-3}T_{\mu \beta }d_{\beta }^{(2)}\right)   \label{MFT1} \\
d_{\beta }^{(2)} &=&\bar{\chi} _{\beta \varepsilon }^{(2)}\left( b_{\varepsilon
}^{(2)}+\lambda R^{-3}T_{\varepsilon \gamma }d_{\gamma }^{(1)}\right) 
\label{MFT2}
\end{eqnarray}%
These equations describe the evolution of each giant spin in an effective
B field containing a time-dependent contribution due to the polarizaton of the other
giant spin. Using (\ref{MFT2}) to eliminate $d_{\beta }^{(2)}$ in(\ref{MFT1}%
), we get 
\[
d_{\alpha }^{(1)}=\bar{\chi} _{\alpha \mu }^{(1)}\left( b_{\mu }^{(1)}+\lambda
R^{-3}T_{\mu \beta }\bar{\chi} _{\beta \varepsilon }^{(2)}\left( b_{\varepsilon
}^{(2)}+\lambda R^{-3}T_{\varepsilon \gamma }d_{\gamma }^{(1)}\right)
\right) 
\]%
\[
\left( \delta _{\alpha \gamma }-\lambda ^{2}\bar{\chi} _{\alpha \mu
}^{(1)}R^{-3}T_{\mu \beta }\bar{\chi} _{\beta \varepsilon
}^{(2)}R^{-3}T_{\varepsilon \gamma }\right) d_{\gamma }^{(1)}=\bar{\chi} _{\alpha
\mu }^{(1)}b_{\mu }^{(1)}+\lambda \bar{\chi} _{\alpha \mu }^{(1)}R^{-3}T_{\mu
\beta }\bar{\chi} _{\beta \varepsilon }^{(2)}h_{\varepsilon }^{(2)}
\]%
Then for $b_{\mu }^{(1)}=0$ (i.e. an external oscillating B field applied only
to moment SPN2) we have 
\[
d_{\alpha }^{(1)}=\chi _{\alpha \varepsilon ,\lambda }^{(12)}b_{\varepsilon
}^{(2)}
\]%
where

\begin{equation}
\bar{\chi} _{\alpha \varepsilon ,\lambda }^{(12)}=\left( 
\varepsilon _{\lambda }^{-1}\right) _{a\beta ,\lambda }^{(12)}S_{\beta
\varepsilon }^{(12)}
\label{ChiEquEpsMinus1S}
\end{equation}
\[
S_{\beta \varepsilon }^{(12)}=\bar{\chi} _{\beta \mu }^{(1)}\lambda R^{-3}T_{\mu
\nu }\bar{\chi} _{\nu \varepsilon }^{(2)}
\]%
\[
\varepsilon _{\lambda }=\left( 
\begin{array}{lll}
1-\lambda ^{2}R^{-6}\bar{\chi} _{1\mu }^{(1)}T_{\mu \beta }\bar{\chi} _{\beta
\varepsilon }^{(2)}T_{\varepsilon 1} & -\lambda ^{2}R^{-6}\bar{\chi} _{1\mu
}^{(1)}T_{\mu \beta }\bar{\chi} _{\beta \varepsilon }^{(2)}T_{\varepsilon 2} & 
-\lambda ^{2}R^{-6}\bar{\chi} _{1\mu }^{(1)}T_{\mu \beta }\bar{\chi} _{\beta \varepsilon
}^{(2)}T_{\varepsilon 3} \\ 
-\lambda ^{2}R^{-6}\bar{\chi} _{2\mu }^{(1)}T_{\mu \beta }\bar{\chi} _{\beta \varepsilon
}^{(2)}T_{\varepsilon 1} & 1-\lambda ^{2}R^{-6}\bar{\chi} _{2\mu }^{(1)}T_{\mu
\beta }\bar{\chi} _{\beta \varepsilon }^{(2)}T_{\varepsilon 2} & -\lambda
^{2}R^{-6}\bar{\chi} _{2\mu }^{(1)}T_{\mu \beta }\bar{\chi} _{\beta \varepsilon
}^{(2)}T_{\varepsilon 3} \\ 
-\lambda ^{2}R^{-6}\bar{\chi} _{3\mu }^{(1)}T_{\mu \beta }\bar{\chi} _{\beta \varepsilon
}^{(2)}T_{\varepsilon 1} & -\lambda ^{2}R^{-6}\bar{\chi} _{3\mu }^{(1)}T_{\mu
\beta }\bar{\chi} _{\beta \varepsilon }^{(2)}T_{\varepsilon 2} & 1-\lambda
^{2}R^{-6}\bar{\chi} _{3\mu }^{(1)}T_{\mu \beta }\bar{\chi} _{\beta \varepsilon
}^{(2)}T_{\varepsilon 3}%
\end{array}%
\right) 
\]%
This becomes simpler if we have strictly uniaxial responses of the individual
spins along (say) the x axis, i.e. 
\[
\bar{\chi} _{\beta \mu }^{(1)}=\delta _{\beta 1}\delta _{\mu 1}\bar{\chi} ^{(1)}
\]%
and similarly for $\bar{\chi} ^{(2)}$. Then we can ignore the 2 and 3 components $%
d_{2}^{(1)},\,d_{3}^{(1)}$ and only need solve a scalar equation, giving 
\begin{equation}
\bar{\chi} _{\lambda }^{(12)}=\frac{\lambda \bar{\chi} ^{(1)}R^{-3}T_{11}\bar{\chi} ^{(2)}}{%
1-\lambda ^{2}R^{-6}\bar{\chi} ^{(1)}T_{11}\bar{\chi} ^{(2)}T_{11}}
\label{ScalarRPAChi12}
\end{equation}%
and then from (\ref{IntEFromChi}) 
\[
E_{\lambda }^{(12)}=-\lambda ^{2}R^{-6}T_{11}^{2}%
\int_{-\infty }^{\infty }\frac{\hbar }{\pi (\exp (\beta \hbar \omega )-1)}%
Im\frac{\bar{\chi} ^{(1)}\bar{\chi} ^{(2)}}{1-\lambda ^{2}R^{-6}\bar{\chi}
^{(1)}T_{11}\bar{\chi} ^{(2)}T_{11}}d\omega 
\label{E12Uniaxial}
\]

From (\ref{HelmholtzFromChi}), the corresponding free energy of the residual interaction is

\begin{eqnarray}
A(R)-A(\infty) =\int_0^1 E^{(12)} \frac{d\lambda}{\lambda} \nonumber\\
=-\frac{1}{2}\int_{-\infty }^{\infty }\frac{\hbar }{\pi (\exp (\beta \hbar \omega )-1)}%
Im \left\{ln( {1-R^{-6}\bar{\chi}
^{(1)}\bar{\chi} ^{(2)}T_{11}^2} ) \right\} d\omega
\label{AUniaxial}
\end{eqnarray}

The corresponding force between SPN1 and SPN2 is 

\begin{equation}
F_\mu = - \frac{\partial A(\vec{R}) }{\partial R_\mu}
\label{forceFromA}
\end{equation}
Eq (\ref{AUniaxial}) is valid for the uniaxial case but is readily generalized: there is in general a sum of logarithms of the eigenvalues of the matrix $1-R^{-6}T\bar{\chi}^{(1)}T\bar{\chi}^{(2)}$.

Note that both $\bar{\chi}^{(1)}$ and $\bar{\chi}^{(2)}$ in  (\ref{ScalarRPAChi12}, 
\ref{AUniaxial}) are frequency-dependent.  If the denominator of (\ref{ScalarRPAChi12}) vanishes for some frequency
$\omega_{0j}$ then we have a finite oscillation of the magnetic moments for zero driving field  - i.e. a free magnon collective oscillation mode of the coupled giant spins. Indeed the free energy (\ref{AUniaxial}) can be related to a sum of the thermal free energies of these magnons.  Actually for the present model, namely  $\bar{\chi}(\omega)=\bar{\chi}_0 / (1-i\omega \tau)$ (see Appendix C) these frequencies will have a large imaginary part (damping), so there are really no magnons in the absence of an applied DC magnetic field. The exception is the case $\omega \approx 0$, where the damping vanishes.  If one of the  magnon frequencies vanishes, $\omega_{0J} =0$, then we have an instability and the system will try to "feeze in" the magnon. This means that the denominator in  (\ref{ScalarRPAChi12}) vanishes for zero frequency, which, as the coupling is increased, will happen first for $\lambda = 1$, i.e.

\begin{equation}
1-R^{-6}\bar{\chi} ^{(1)}(0) \bar{\chi} ^{(2)}(0)T_{11}^2 = 0 .
\label{RPAInstabilityCondition}
\end{equation}
In time-dependent mean-field theories such as this, this behaviour is usually taken to indicate a transition to a broken-symmetry state - in this case the moments presumably freeze into a permanent ordering in the antiparallel configuration.

\section{Energy in second order (weak coupling)}

Note that if we only want the energy to second order in the interaction then
from (\ref{IntEFromChi}) we only need $\bar{\chi} ^{(12)}\,$ to first order in $T_{ij}$, so we can take $\varepsilon _{\lambda }=I$ in (\ref{ChiEquEpsMinus1S}),  giving 
\begin{equation}
\bar{\chi} _{\alpha \varepsilon \lambda }^{(12)}\approx \lambda \bar{\chi} _{\beta \mu
}^{(1)}R^{-3}T_{\mu \nu }\bar{\chi} _{\nu \varepsilon }^{(2)} .
\label{Chi12toFirstOrder}
\end{equation}%
so that (\ref{IntEFromChi}) becomes, since $\int_{0}^{1}\lambda ^{2}\frac{%
d\lambda }{\lambda }=\frac{1}{2}$, 
\begin{eqnarray}
A^{residual} &=&A(R)-A(\infty )=\int_{0}^{1}E_{\lambda }^{(12)}\frac{%
d\lambda }{\lambda }\\
&=&-\frac{1}{2}R^{-6}\sum_{i,j,\mu ,\nu
=1}^{3}T_{ij}(\hat{R})I_{ij\mu \nu },\;\;where  \label{SecondOrderEnergy} \\
I_{ij\mu \nu }=\int_{-\infty }^{\infty }&\frac{\hbar }{\pi
(\exp (\beta \hbar \omega )-1)}&Im \left[ \bar{\chi} _{i\mu }^{(1)}(\omega
+i0)T_{\mu \nu }(\hat{R})\bar{\chi} _{\nu j}^{(2)}(\omega +i0)\right]d\omega .
\end{eqnarray}%
The $R^{-6}$ dependence is apparent. \ In the case of uniaxial
response 
\begin{eqnarray}
A^{residual} &=&-\frac{1}{2}R^{-6}T_{11}(\hat{R})T_{11}(\hat{R})I,\;\; where
\label{AWeakUniaxial} \\
I =\int_{-\infty }^{\infty }\frac{\hbar }{\pi (\exp (\beta
\hbar \omega )-1)}&Im&\left[ \bar{\chi} ^{(1)}(\omega +i0)\bar{\chi} ^{(2)}(\omega
+i0)\right] d\omega 
\end{eqnarray}

From Appendix C, a simple model for a
superparamagnetic susceptibility is $\bar{\chi}(\omega) =\bar{\chi} _{0}/(1-i\omega \tau ) $, and
the frequency integral $I$ in (\ref{AWeakUniaxial}) can be estimated
analytically in two limits depending on the thermal flipping time $\tau $ of
the giant spins (see Eqs (\ref{FrequIntegralQuantal}) and (\ref{FrequIntegralClassical}) 
of Appendix C). This gives a residual free energy

\begin{equation}
A^{residual}=-\frac{1}{2}R^{-6}T_{11}^{2}\bar{\chi} _{0}^{2}\left\{
\begin{array}{l}
k_{B}T,\;\;\;\;k_{B}T>>\hbar \tau ^{-1} \\ 
\frac{1}{\pi }\hbar \tau ^{-1},\;\;k_{B}T<<\hbar \tau ^{-1}
\end{array} 
\right.
\label{AweakUniaxial2Limits}
\end{equation}
and a residual van-der-Waals-like force 

\begin{eqnarray}
F^{residual}= -\frac{\partial A^{residual}}{\partial R}
 = 3R^{-7}T_{11}^{2}\bar{\chi} _{0}^{2}\left\{
\begin{array}{l}
k_{B}T,\;\;\;\;k_{B}T>>\hbar \tau ^{-1} \\ 
\frac{1}{\pi }\hbar \tau ^{-1},\;\;k_{B}T<<\hbar \tau ^{-1} %
\end{array}%
\right.
   \label{FweakUniaxial2Limits}
\end{eqnarray}

\section{Orders of magnitude}

\subsection{\protect\bigskip SPNs below the blocking temperature}
\label{SPNsBelowTb}
First consider the energy and force of interaction between two SPNs \emph{%
below }their blocking temperature so that each has a permanant magnetic
moment of magnitude $d_{0}=n\mu _{B}$. At separation $R$ the direct
dipole-dipole energy is dependent on orientation but is of order%
\begin{eqnarray}
\left| E^{direct}\right| \approx \frac{\mu _{0}}{4\pi }%
d_{0}^{2}R^{-3}=(10^{-7})(9\times 10^{-24}n)^{2}(10^{9})^{3}\left[ R/10^{-9}%
\right] ^{-3} \nonumber \\
=8.1\times 10^{-27}n^{2}\left( \frac{10^{-9}\;m}{R}\right) ^{3}\;\;Joule  
\label{Edirect}
\end{eqnarray}%
For example if $n=1000$ and $R=1\;nm$,\ $E^{direct}\approx 10^{-20}J.$ At $%
T=300K$ the thermal energy is $k_{B}T_{room}=4\times 10^{-21}J\;$, so $%
E^{direct}\approx 2k_{B}T$. Thus if the two SPNs are not thermally suppressed at $%
T=300K$ and are able to approach to within a nanometer, they will not be
prevented by thermal effects from rotating to the antiparallel configuration
and binding (clumping).

The corresponding force $F^{direct}$ between the SPNs is highly
orientation-dependent but is of order%
\begin{eqnarray}
\left| F^{direct}\right| \approx \frac{\mu _{0}}{4\pi }3d_{0}^{2}R^{-4}%
\approx 3\times 8.\,\allowbreak 1\times 10^{-27}(10^{-9})^{-1}n^{2}\left( 
\frac{10^{-9}\;m}{R}\right) ^{4}\;N  \nonumber \\
=2\times 10^{-17}n^{2}\left( \frac{%
10^{-9}\;m}{R}\right) ^{4}\;N.  \label{Fdirect}
\end{eqnarray}%
For $n=1000$ and $R=1\;nm$ this gives a force of order $20pN$, which is
small but should be directly detectable via Atomic Force Microscopy (AFM)
with single SPNs attached to substrate and tip.

\subsubsection{SPNs above the blocking temperature}
\label{SectionSPNsAboveTb}

Now consider a similar system but with a blocking temperature below room
temperature so that at $300K$ there are no permanent moments. Then the 
vdW-like theory derived above gives the free energy of interaction. \ For
numerical estimates we assume uniaxial susceptibilites and work in the
weak-coupling limit. We also assume that the giant spins have a
zero-frequency susceptibility 
\begin{equation}
\bar{\chi} _{0}=\frac{\left( n\mu _{B}\right) ^{2}}{k_{B}T}
\label{ModelForChi0}
\end{equation}%
corresponding to a giant moment of $n\,$\ Bohr magnetons.  Then (\ref%
{AweakUniaxial2Limits}) gives 
\begin{equation}
A^{residual}\approx \left\{ 
\begin{array}{cc}
-2 \times 10^{-7}\frac{n^{4}}{T^{2}}\left( \frac{10^{-9}\; m}{R}%
\right) ^{6} \hbar \tau^{-1}\,\,Joule, & \hbar \tau ^{-1}>>k_{B}T \\ 
-2 \times 10^{-7}\frac{n^{4}}{T^{2}}\left( \frac{10^{-9}\; m}{R}\right)
^{6}\,k_{B}T\;Joule, & \hbar \tau ^{-1}<<k_{B}T%
\end{array}%
\right.   \label{Eresidual}
\end{equation}%
For example, let $n=1000$, $R=1\;nm$ , $T=300K$, and $\hbar \tau
^{-1}<<k_{B}T\ $. $\ $Then $A^{residual}\approx -2 \times 10^{-7}\frac{(1000)^{4}}{%
(300)^{2}}k_{B}T$ $=2k_{B}T.$ This means that for the present case the
residual energy predicted by the perturbative theory is about the same as the
direct energy (\ref{Edirect}), which is unphysical and simply means that the weak-coupling
condition is not met and we need (at least) the full RPA theory here (Eq. (\ref{AUniaxial})). \ If
we are in the limit $\hbar \tau ^{-1}>>k_{B}T$ the residual interaction will
be even larger. \ In this case the system of two SPNs, despite the thermal
averaging of an individual SPN, is most probably near to a trasition to a
spin-locked configuration. \ In the RPA theory the onset of this condition
would correspond to a zero denominator in (\ref{ScalarRPAChi12}). \ This
would occur for $R^{-6}\bar{\chi}_0 ^{(1)}T_{11}\bar{\chi}_0 ^{(2)}T_{11}\approx 1\,$, i.e. 
\begin{eqnarray*}
R_{lock} &\approx &(T_{11}\bar{\chi} _{0})^{1/3}
\approx \left( \frac{10^{-7}(n\mu
{B})^{2}}{k_{B}T}\right) ^{1/3}
\approx
10^{-2}\left( \frac{300}{T}\right) ^{1/3}n^{2/3}\;nm
\end{eqnarray*}
For the present case with $n=1000$ and $T=300K$, the crossover occurs at
about $R_{lock}=1\;nm$, which is consistent with the above finding that the
perturbative calculation of the attraction at this separation was unphysical. 

To give another example, suppose that $n=100$, $R=10nm$, $T=300K$, and $%
\hbar \tau ^{-1}<<k_{B}T$ .Then $A^{residual}\approx -2 \times 10^{-6}\frac{%
(100)^{4}}{(300)^{2}}(\frac{1}{10})^{6}k_{B}T=2 \times 10^{-10}k_{B}T,$whereas the
direct interaction between permanent moments under the same conditions from
(\ref{Edirect}) is $E^{direct}=8.1\times 10^{-27}100^{2}\left( \frac{1}{10}%
\right) ^{3}/(300(1.24\times 10^{-23}))k_{B}T=2\times 10^{-5}k_{B}T.$ So for
this example, neither the direct nor the residual interaction would tend to
lock the SPNs into an antiferromagnetically aligned pair. The mechanical
forces on the SPNs due to the spin-spin interaction in either the direct or
the thermally smeared residual case would be negligible in the context of normal
Brownian motion.

\section{Clumping considerations in fluid suspension of SPNs}

(i) Consider SPNs with $n=1000$ and with a bocking temperature satisfying $%
T_{b}>300K$ . From the numbers shown above, at $T=300K$, if they are able to
aproach one another within about a nanometer, these particles will form
pairs or larger clusters (''clumping'') that are due to the direct
(not thermally suppressed) magnetic dipolar interaction, and that are not
readily broken by thermal processes.  Furthermore under these same conditions the 
demagnetization field inside a single SPN might be significant, so that the SPN would 
no longer contain a single domain as assumed so far.\ At larger separations $R$ the binding
energy falls off as $R^{-3}$, and so the direct magnetic energy, as $R$ is
increased, will soon be less than the thermal energy $k_{B}T$. \ The
interaction at these larger separations will not immediately cause binding,
but may well determine the kinetics of closer approach between
nanoparticles, resulting ultimately in clumping when shorter separations are
attained. \ This process is complicated by the strong orientational
dependence of the direct interaction (\ref{InteractionBetwMagnets}). SPNs
will tend to rotate mechanically within in the fluid, in order to minimize
the free energy in the ''antiferromagnetic'' relative orientation, after
which their mutual force is attractive. \ Thus the kinetics of clumping will
be far from straightforward.

If clumping is undesirable, the $n^{2}R^{-3}$ dependence of of the direct
SPN-SPN magnetic binding energy suggests that smaller SPNs (e.g. $n=100$)
will be desirable because they are less susceptible to clumping, i.e. they
can approach to smaller distances (e.g. $R=\sqrt[3]{10^{-2}}\,nm=$ $0.2$ $nm$%
) before clumping occurs. $\ $\ In fact, at such small separations $R$, the
point dipole approximation used here may break down, softening the
interaction and possibly leading to the conclusion that the binding energy
even at contact is less than the thermal energy. This would imply minimal
clumping.

(ii) Consider SPNs in suspension at $T=300K$, with $n=1000$, but now with $T_{b}<300K.$
Here, despite the thermal suppression of the net individual moments, there is
a uniformly attractive residual magnetic SPN-SPN free energy  $A^{residual}$
. This varies as $n^{4}R^{-6}$ within the perturbative approximation (see (%
\ref{Eresidual})), and so becomes much weaker than the direct interaction at
large separations $R$ . However at shorter separations, the stronger $n$ and 
$R$ dependence of the perturbative residual energy expression (\ref%
{Eresidual}) suggests that $E^{resid}$ could exceed $E^{direct}.$ This is of
course unphysical: the correlations between the orientations of giant
moments that give rise to $E^{resid}$ cannot be greater than perfect
correlation, corresponding to the direct interaction in the
antiferromagnetic configuration \ of the two giant moments. Thus in general $%
\left| E^{resid}\right| \leq \left| E^{direct,\max }\right| .$ In fact the
perturbative approximation breaks down in small-$R$ regime, and the full RPA
expression (\ref{AUniaxial}) will be needed
instead of (\ref{Eresidual}). We do not yet have analytic energy and force
expressions in this regime. \ However it is clear that this approach can
yield a residual interaction $\left| E^{resid}\right| $ of a strength
approaching $\left| E^{direct,\max }\right| $. It seems likely, therefore,
that because of the residual interaction, there will not always be a discontiunous
cessation of clumpimg as the temperature is raised above the blocking
temeparure $T_{b}$.  However the direct interaction can be repulsive whereas the residual 
interaction is always attractive, so there is scope for some quite rich behaviour.

\section{Prospects for experimental verification of the theory}

\subsection{Direct measurement of the force between two individual SPNs}

In Section \ref{SPNsBelowTb} above, the direct interaction between permanently magnetized SPNs with $%
n=1000$ at separation $R=1\,nm$ was estimated to exceed $k_{B}T_{room}$, and
the force was estimated as $20\,pN$. A force of this magnitude is likely to
be observable, with some care, via atomic Force Microscopy. The simplest
configuration might involve one SPN attached to a non-magnetic substrate,
and another SPN attached to the AFM tip. \ \ One could then measure the
force as a function of temperature. \ \ One might expect a reduction in the
measured force as $T$ is increased above $T_{b}$.  As discussed above, the 
force could even change from repulsive to attractive, depending on the initial
 orientation of the giant moments prior to heating and subsequent destruction of 
the net moments. \ The need for a
measurably large force puts us out of the perturbative regime for the
residual interaction, so more straightforward but messy theoretical work
will be required in order to predict the way in which $F$ varies with
distance and temperature near $(R,T)=(1nm,T_{b})$. \ It is not clear whether
the force will be large enough for AFM\ detection in the regime of larger
separations where the perturbative analysis (\ref{Eresidual}) is valid.

\subsection{Indirect measurement via observation of structure factors in
fluid suspension}

Here we propose (e.g.) small-angle xray diffraction measurements on SPNs in
suspension in a viscous fluid such as glycerine. The metallic SPNs should
provide good Xray contrast. The measured structure factor of the array of
SPNs should reveal evidence of positional corrrelations between the SPNs,
which in turn is related to the forces between the SPNs as predicted here.
Again, one hopes to see some changes as the temperature is raised through
the blocking temperature $T_{b}$.

\subsection{Magnetic resonance experiments}

Although the present theory did not predict any lightly damped magnons 
(combined oscillations of the magnetic moments) for a pair of adjacent SPNs, 
there might be the possibility of such modes if a strong DC magnetic field is
 applied.  Magnetic resonance experiments might then be able to detect shifts 
in the single-SPN resonance frequency due to the proximity of a pair of SPNs. 
 Even without the external DC field, an analysis of the linewidth of the
 zero-frequency "resonance" might reveal information about SPN-SPN coupling.

\section{Summary and future directions}

We have predicted a residual force between superparamagnetic nanoparticles
that persists above the blocking temperature. The force is the magnetic
analogue of the electrically-driven van der Waals interaction between
electrically neutral molecules. Our theory also deals with the dynamic spin
response of coupled SPNs to small ac external magnetic fields. \ Our results
may be experimentally testable, and may have implications for ferro-fluids, for 
nanoparticle-based medical therapies, and for magnetic recording technology.
The new force is most likely to be significant for nanoparticles that approach
one another quite closely, at separations of O(nm).  At these separations the 
point-magnetic-dipole approximation used here will need to be replaced by a theory 
that attributes a finite spatial size and definite physical shape to the nanoparticles.
A good starting model will be an ellipsoidal shape, and fortunately the full electrodynamic theory of Casimir interactions is quite well developed for this 
geometry.  A theory along these lines will be the next step.

\section{Appendix A: How to deal with the entropic part (classical
angle-distribution approach)}

The joint state of two interacting superparamagnets is specified by a
classical distribution $f^{(2)}(\Omega _{1},\Omega _{2})$ in the 
two solid angles $\Omega _{1},\Omega _{2}\,$%
defining spatial directions where the 2 giant spins point: 

The reduced-strength interaction $\lambda E$ between the superparamagnets is
given by (\ref{E12Lambda}). Then from general
thermodynamic principles, at a given
temperature $T$ ,coupling strength $\lambda $ and separation $R$, the
correct distribution $f_{\lambda }^{(2)}\,(T,R)$ is that which minimizes the
trial free energy:
\[
A(\lambda ,T,R)=Min_{f^{(2)}:\left| f^{(2)}\right| =1}A(\lambda
,T,R:[f^{(2)}])
\]

so that the following functional derivative is zero%
\begin{equation}
0=\frac{\delta A}{\delta f^{(2)}},\;\;where  \label{FuncDerivOfFVanishes}
\end{equation}%
\begin{eqnarray*}
A(\lambda ,T,R :[f^{(2)}])=\lambda E-TS \\
=\lambda R^{-3}\sum_{ij=1}^{3}T_{ij}(\hat{R})\sum_{\Omega _{1},\Omega
_{2}}d_{i}^{(1)}(\Omega _{1})d_{j}^{(2)}(\Omega _{2})f^{(2)}(\Omega
_{1},\Omega _{2}) \\
+k_{B}T\sum_{\Omega _{1},\Omega _{2}}f^{(2)}(\Omega
_{1},\Omega _{2})\ln f^{(2)}(\Omega _{1},\Omega _{2})
\end{eqnarray*}%
Consider an infinitesimal increase in the coupling strength from $\lambda \,$%
to $\lambda +\Delta \lambda $. As a result, $f^{(2)}\,$changes by an amount $%
\Delta f^{(2)}$ and noting that $E=<H>_{\lambda} = E_{\lambda}/\lambda$ we have a resulting change in $A:$ 
\[
\Delta A=\Delta \lambda E+\sum_{\Omega _{1},\Omega _{2}}\frac{%
\delta A}{\delta f^{(2)}}\Delta f^{(2)}=\Delta \lambda \frac{E_{\lambda }}{\lambda}+0 
\]%
where the zero comes from (\ref{FuncDerivOfFVanishes}).

Notice that we only have to know the interaction $E\,$and not the entropic
part, to find the change in $A$.

Then the change in $A$ in switching on the interaction adiabatically is 
\[
A_{\lambda =1}-A_{\lambda =0}=\int_{0}^{1}\Delta A=\int_{0}^{1}E_{\lambda
}\frac{d\lambda}{\lambda} 
\]%
We have already shown how to calculate $E_{\lambda }$ by using the
fluctuation-dissipation theorem and the mean-field (RPA) assumption. Also
note that the $\lambda =0$ value of the free energy is independent of
separation $R$: 
\begin{eqnarray*}
f_{\lambda =0}^{(2)}(\Omega _{1},\Omega _{2}) &=&f_{0}^{(a)}(\Omega
_{1})f_{0}^{(b)}(\Omega _{2}), \\
A(\lambda &=&0,T,R)=k_{B}T\sum_{\Omega _{1},\Omega _{2}}f_{\lambda
=0}^{(2)}(\Omega _{1},\Omega _{2})\ln f_{\lambda =0}^{(2)}(\Omega
_{1},\Omega _{2}) \\
&=&k_{B}T\left( \sum_{\Omega _{1}}f_{\lambda =0}^{(a)}(\Omega _{1})\ln
f_{\lambda =0}^{(a)}(\Omega _{1})+\sum_{\Omega _{2}}f_{\lambda
=0}^{(b)}(\Omega _{2})\ln f_{\lambda =0}^{(b)}(\Omega _{2})\right) .
\end{eqnarray*}%
Thus the entire $R\,$dependence of $A(\lambda =1,R,T)$ is captured by the
integral $\int_{0}^{1}E_{\lambda }\frac{d\lambda}{\lambda} .$

\section{Appendix B: How to deal with the entropic part (fully quantal
approach)}

Our quantum mechanical basis (NOT the eigenstates) for the combined magnetic
state of the two systems together consists of the factorised states%
\[
\left| ij\right\rangle =\left| i\right\rangle \left| j\right\rangle \;. 
\]%
where the first ket refers to\ quantum state of SPN1 and the second ket to
SPN2.

The thermal density matrix operator of a pair of magnetically interacting
nanoparticles has matrix elements in this basis denoted by 
\[
\rho _{ij:kl}^{(2)} 
\]%
and traces can be taken over this or any other basis with the same result.

We consider starting from the thermal equilibriumn of two isolated
nanoparticles, and consider the effect on the free energy of turning on the
interaction by replacing the inter-nanoparticle interaction hamiltonian $%
\hat{H}^{(12)}(R)$ by $\lambda \hat{H}^{(12)}(R)$, and then increasing $%
\lambda $ from $0$ to $1$ while holding the inter-particle separation $R$
fixed.

For coupling strength $\lambda $ the Helmholtz free energy is a trace:%
\[
A(\lambda ,T,R,[\hat{\rho}^{(2)}])=E-TS=Tr\left( \left( \hat{H}_{0}+\lambda 
\hat{H}^{(12)}(R)\right) \hat{\rho}^{(2)}\right) -k_{B}Tr\left( \hat{\rho}%
^{(2)}\ln \hat{\rho}^{(2)}\right) 
\]%
For fixed Hamiltonian, at thermal equilibrium $A$ is stationary with respect
to arbitrary variations in density matrix that preserve $Tr\hat{\rho}^{(2)}$%
( see e.g. []):

\[
A(\lambda ,T,R)=Min_{\rho ^{(2)}:tr\rho ^{(2)}=1}A(\lambda ,T,R:[\hat{\rho}%
^{(2)}])
\]%
so that%
\begin{equation}
0=\frac{\delta A}{\delta \hat{\rho}^{(2)}}\;.
\end{equation}%
Then the first-order change in the equilibrium free energy, when the
coupling is increased from $\lambda $ to $\Delta \lambda $, is%
\[
\Delta A=\Delta \lambda Tr\left( \hat{H}^{(12)}(R)\hat{\rho}_{\lambda
}\right) +\lambda Tr\left( \frac{\delta A}{\delta \hat{\rho}^{(2)}}\Delta 
\hat{\rho}^{(2)}\right) =\Delta \lambda \frac{E_{\lambda }^{(12)}}{\lambda}+0
\]%
Then the change in free energy in switching on the interaction between the
two systems is
\begin{equation}
A(\lambda =1,T,R)-A(\lambda =0,T,R)=\int_{0}^{1} E_{\lambda
}^{(12)} \frac{d\lambda}{\lambda} 
\end{equation}

The same formula can be derived for the classical case, by considering a
pair distribution $f(\Omega _{1},\Omega _{2})$ of angular orientations $%
\Omega $ of the two giant moments.  (See Appendix A).

\section{Appendix C: simple superparamagnetic model for $\protect\bar{\chi} _{%
\protect\alpha \protect\mu }(\protect\omega )$}

Assume that the individual giant moment has its easy axis along $\hat{e}$. 
Then the response to a field $\vec{h}$ is only via $\vec{h}.\hat{e}$, and
the response is along $\hat{e}$

i.e.\thinspace 
\[
\vec{d}=\bar{\chi} (\omega )(\vec{b}.\hat{e})\hat{e}
\]%
\[
\bar{\chi} _{\alpha \mu }(\omega )=\hat{e}_{\alpha }\hat{e}_{\mu }\bar{\chi} (\omega )
\]
Here we assume a widely-used model \cite{ThFluctSPNFullerBrown} for the frequency-dependent magnetic susceptibility of an individual SPN in the absence of a d.c. external magnetic field:
\[
\bar{\chi} (\omega )=\frac{\bar{\chi} _{0}}{1-i\omega \tau }\,\,,\,\,\tau =\tau _{0}\exp
(E_{0}/(k_{B}T))\,\,\,
\]%
where $\tau $ is the thermal flipping time of the giant moment, assumed to
arise from an intrinsic attempt time $\tau _{0}$ and a thermally-activated
Boltzmann success rate in surmounting the  anosotropy energy barrier $E_{0}$.
Then 
\begin{eqnarray*}
Im\left[ \bar{\chi} ^{(1)}(\omega +i0)\bar{\chi} ^{(2)}(\omega +i0)\right]  &=&%
Im\left[ \frac{\bar{\chi} _{0}^{(1)}}{1-i\omega \tau ^{(1)}}\frac{\bar{\chi}
_{0}^{(2)}}{1-i\omega \tau ^{(2)}}\right]  \\
&=&Im\left[ \frac{\bar{\chi} _{0}^{(1)}(1+i\omega \tau ^{(1)})}{1+\omega
^{2}\tau ^{(1)2}}\frac{\bar{\chi} _{0}^{(2)}(1+i\omega \tau ^{(2)})}{1+\omega
^{2}\tau ^{(2)2}}\right]  \\
&=&\frac{\bar{\chi} _{0}^{(1)}\bar{\chi} _{0}^{(2)}\omega (\tau ^{(1)}+\tau ^{(2)})}{%
\left( 1+\omega ^{2}\tau ^{(1)2}\right) \left( 1+\omega ^{2}\tau
^{(2)2}\right) }
\end{eqnarray*}%
If the two superparamagnetic nanoparticles have the same parameters, we get 
\begin{equation}
Im\left[ \bar{\chi} ^{(1)}(\omega +i0)\bar{\chi} ^{(2)}(\omega +i0)\right] =\frac{%
2\bar{\chi} _{0}^{2}\omega \tau }{\left( 1+\omega ^{2}\tau ^{2}\right) ^{2}}
\label{ImChiChi}
\end{equation}

Then the second order energy (\ref{SecondOrderEnergy}) between two
nanoparticles with easy axes $\hat{e}^{(1)}$, $\hat{e}^{(2)}$ becomes 
\begin{eqnarray*}
<E^{12}{}_{\lambda }>\;=\lambda ^{2}R^{-3}\sum_{ij=1}^{3}T_{ij}(\hat{R})%
\int_{-\infty }^{\infty }\frac{\hbar }{\pi (\exp (\beta \hbar
\omega )-1)} \\
\times Im\bar{\chi} _{i\mu }^{(1)}(\omega +i0)R^{-3}T_{\mu \nu }(%
\hat{R})\bar{\chi} _{\nu j}^{(2)}(\omega +i0)d\omega  \\
\\
=\lambda ^{2}R^{-6}\sum_{ij=1}^{3}T_{ij}(\hat{R})\hat{e}_{i}^{(1)}\hat{e}%
_{\mu }^{(1)}T_{\mu \nu }(\hat{R})\hat{e}_{\nu }^{(2)}\hat{e}_{j}^{(2)} \\
\times \int_{-\infty }^{\infty }\frac{\hbar }{\pi (\exp (\beta \hbar
\omega )-1)}\bar{\chi} ^{(1)}(\omega +i0)\bar{\chi} ^{(2)}(\omega +i0)d\omega
\end{eqnarray*}%
\ For two similar SPNs the frequency integral is 
\begin{equation}
I=\int_{-\infty }^{\infty }\frac{\hbar }{\pi (\exp (\beta
\hbar \omega )-1)}\frac{2\bar{\chi} _{0}^{2}\omega \tau }{\left( 1+\omega ^{2}\tau
^{2}\right) ^{2}}d\omega   \label{FrequInt2ndOrder}
\end{equation}%
If $\beta \hbar /\tau <<1\,$then the $Im\bar{\chi} ^{2}$ factor cuts the
integral off for $\left| \omega \right| >\tau ^{-1}\,$, i.e. for 
$\beta \hbar \omega > 1$ and we can
Taylor-expand the denominator to 1st order giving 
\begin{eqnarray}
I &=&\int_{-\infty }^{\infty }\frac{\hbar }{\pi \beta \hbar
\omega }\frac{2\bar{\chi} _{0}^{2}\omega \tau }{\left( 1+\omega ^{2}\tau
^{2}\right) ^{2}}d\omega =k_{B}T\frac{2\bar{\chi} _{0}^{2}}{\pi}\tau \left(
\int_{-\infty }^{\infty }\frac{1}{(1+\omega ^{2}\left| \tau \right| ^{2})^{2}%
}d\omega \right) \nonumber \\
&=&k_{B}T\frac{2\bar{\chi} _{0}^{2}}{\pi }\tau \left( \frac{1}{2}\pi \frac{%
\tau }{\tau ^{2}}\right) =\bar{\chi} _{0}^{2}k_{B}T \label{FrequIntegralClassical}
\end{eqnarray}%
On the other hand if $\beta \hbar /\tau >>1$ then 
\begin{eqnarray}
I\approx \frac{\hbar }{\pi}\int_{-\infty }^{0}\frac{-2\bar{\chi}
_{0}^{2}\omega \tau }{\left( 1+\omega ^{2}\tau ^{2}\right) ^{2}}d\omega =-%
\frac{2\hbar \bar{\chi} _{0}^{2}\tau }{\pi}\int_{-\infty }^{0}\frac{\omega }{%
\left( 1+\omega ^{2}\left| \tau \right| ^{2}\right) ^{2}}d\omega  \nonumber \\
 =-\frac{%
2\hbar \bar{\chi} _{0}^{2}\tau }{\pi}\left( -\frac{1}{2\tau ^{2}}\right) =%
\bar{\chi} _{0}^{2}\frac{\hbar }{\pi }\tau ^{-1}  \label{FrequIntegralQuantal}
\end{eqnarray}


\end{document}